\documentclass{article}
\usepackage[utf8]{inputenc}
\usepackage{amsmath,amssymb}
\usepackage{appendix}
\usepackage{slashed}
\usepackage{comment}
\usepackage{authblk}
\usepackage{xcolor}

\newcommand{\bs}[1]{\boldsymbol{#1}}
\newcommand{\nn}{\nonumber}
\newcommand{\<}{\langle}
\renewcommand{\>}{\rangle}

\evensidemargin=\oddsidemargin
\title{\bf From chiral to conformal anomalies: a double copy perspective for CFT correlators}
\author[a,b,c]{Claudio Corianò}
\author[a,b]{Stefano Lionetti}

\affil[a]{Dipartimento di Matematica e Fisica, Universit\`{a} del Salento 
	and INFN Sezione di Lecce,Via Arnesano 73100 Lecce, Italy}
\affil[b]{National Center for HPC, Big Data and Quantum Computing}
\affil[c]{CNR-Nanotec, Institute of Nanotechnology, Lecce, Italy}
\date{}

\begin{document}

	\maketitle
	\begin{abstract}
		
		We uncover a double-copy relation between chiral and conformal anomalies in four-dimensional conformal field theory. We show that this relation extends from local anomaly structures to complete momentum-space three-point correlators and, through their flat-space limits, to five-dimensional scattering amplitudes. In particular, the parity-odd current correlators associated with the chiral anomaly reduce, in the flat-space limit, to a universal amplitude generated by a Chern--Simons interaction. Squaring these anomalous current correlators reproduces the contribution to the stress-tensor three-point function $\langle TTT\rangle$, controlled by the conformal anomaly, thereby establishing a direct link between chiral- and conformal-anomaly structures.
	\end{abstract}

\newpage

\section{Introduction}

The double copy has revealed one of the most striking connections in modern quantum field theory: under suitable conditions, gravitational interactions can be reconstructed from products of gauge-theory structures \cite{Kawai:1985xq,Bern:2008qj,Bern:2010ue}. First identified in the relation between Einstein gravity and pure Yang--Mills theory, the double copy has since been extended to a broad class of theories, including higher-derivative conformal gravity, higher-derivative gauge theories and bi-adjoint scalar theories \cite{Broedel:2012rc,Johansson:2017srf,Johansson:2018ues}. It has also been generalized to higher-point and loop amplitudes \cite{Bern:2013yya,He:2017spx}. For a comprehensive review, see \cite{Bern:2019prr}.
\\
Originally discovered in the context of scattering amplitudes, the double copy has since appeared in a wide range of settings, suggesting that it may reflect a more general principle extending beyond its original formulation. Momentum-space conformal field theory provides a natural arena in which to explore this possibility.
\\
Conformal correlators are strongly constrained by symmetry, yet retain a rich tensorial structure that can encode information usually associated with scattering amplitudes. In particular, scattering amplitudes can be extracted from suitable limits of CFT correlators in position, momentum and Mellin space \cite{Gary:2009ae,Gary:2009mi,Komatsu:2020sag,Penedones:2010ue,Raju:2012zr,Marotta:2024sce}, providing a CFT derivation of a variety of flat-space amplitude results \cite{Fitzpatrick:2011hu,Fitzpatrick:2011dm}. The connection also works in the opposite direction, with amplitude-based methods having recently been applied to the study of CFT correlators \cite{Caron-Huot:2017vep,Gillioz:2020mdd}. This interplay suggests that double-copy relations among amplitudes should leave a precise imprint on the structure of conformal correlators.
\\
Momentum-space conformal correlators have become an extensively studied subject in recent years \cite{Coriano:2013jba,Bzowski:2013sza}. Such correlators are particularly interesting in the presence of quantum anomalies \cite{Bzowski:2015pba,Bzowski:2017poo,Bzowski:2018fql,Coriano:2023hts,Coriano:2023cvf,Coriano:2023gxa}.\\
Quantum anomalies provide a remarkable window into the interplay between symmetries, quantum effects, and topology. They arise when a symmetry of the classical theory fails to survive quantisation. Among the most familiar examples are the chiral and conformal anomalies. The former manifests itself through the anomalous non-conservation of an axial current, while the latter appears as a non-vanishing trace of the stress-energy tensor. In four dimensions, both anomalies contain contributions involving topological densities, suggesting a deeper connection between structures that, at first sight, belong to distinct sectors of a quantum field theory.
\\
An intriguing relation between these two classes of anomalies was identified in \cite{Bzowski:2017poo}, where the tensor structures associated with the Euler density and the gauge-field Pontryagin density were shown to obey a double-copy-like factorisation. More precisely, the gravitational structure associated with the type-A conformal anomaly can be expressed as a product of two gauge-theory structures associated with the chiral anomaly. In the context of conformal field theory, however, the relation established at the level of local anomaly densities leaves open a natural question: does such a factorisation extend to the full correlation functions governed by these anomalies?
\\
In this work, we investigate this question in four-dimensional conformal field theory. Our main result is that the double-copy relation between chiral and conformal anomalies extends beyond their local anomaly densities to complete momentum-space three-point correlators. The connection becomes particularly transparent when the correlators are uplifted to one higher dimension and their flat-space limits are taken. In this framework, the singular behaviour of four-dimensional CFT correlators in the flat-space limit encodes scattering amplitudes in five-dimensional Minkowski space. This provides a natural setting in which the double-copy structure can be formulated directly in terms of gauge- and gravity-theory amplitudes, making the relation between chiral- and conformal-anomaly structures manifest.
\\
Double-copy relations for conformal correlators have been explored in several contexts. In particular, \cite{Farrow:2018yni,Lipstein:2019mpu} established double-copy relations for the parity-even sector of CFT correlators. The parity-odd sector was subsequently studied in \cite{Jain:2021qcl}; however, that analysis focused on three-dimensional correlators, where anomalies do not arise. Our analysis extends these double-copy relations to the parity-odd sector of four-dimensional CFT correlators in the presence of anomalies.
\\
The paper is organised as follows. In Section~\ref{sec:dcopyanomaly}, we review the relation between the gauge and gravitational anomaly densities identified in \cite{Bzowski:2017poo} and derive a new analogous relation involving mixed parity.
\\
Section~\ref{sec:dcopyttojjo} provides a first exploration of double-copy relations at the level of CFT correlators. In particular, we consider a class of simple correlators involving a marginal scalar operator and either two conserved currents or two stress tensors, namely $\langle JJO\rangle$ and $\langle TTO\rangle$. By separating these correlators into parity-even and parity-odd sectors, we identify double-copy relations in which products of gauge-theory structures reproduce the corresponding gravitational structures.
\\
In Section~\ref{review}, we review the uplift of momentum-space CFT correlators to higher-dimensional flat-space amplitudes. We also introduce the 3K integrals that naturally arise in the momentum-space representation of conformal correlators and review their behaviour in the flat-space limit.
\\
In Section~\ref{sec:chiranomalcorrel}, we turn to the chiral-anomaly correlators. We review the determination of the three-point functions with two vector and one axial-current insertions, $\langle J_V J_V J_A\rangle$, and with three axial-current insertions, $\langle J_A J_A J_A\rangle$, by imposing the conformal constraints. We then determine their flat-space limits, which reduce to a five-dimensional parity-odd amplitude generated by a Chern--Simons interaction. An important subtlety concerns Schouten identities: in four dimensions, different parity-odd tensor structures can represent the same correlator because of dimension-dependent identities. This allows the correlators to be written in apparently different forms, some of which may obscure double-copy relations while others make them manifest.
\\
The gravitational side is provided by the parity-even three-point function of the stress-energy tensor, $\langle TTT\rangle$, discussed in Section~\ref{sec:confanomalcorrel}. We review the solution of the conformal constraints for this correlator and its flat-space limit, which contains several contributions associated with different gravitational interactions. We isolate the part controlled by the trace-anomaly coefficients, which is the relevant contribution for the double-copy relation.
\\
In Section~\ref{sec:dcopyavvttt}, we combine these results to establish the double-copy relation between the chiral- and conformal-anomaly correlators.
Finally, in Section~\ref{sec:conclusions}, we summarise our main results and briefly comment on possible directions for extending this analysis.

\section{Double copy relations between anomaly densities}\label{sec:dcopyanomaly}

The starting point of our analysis is the relation outlined in \cite{Bzowski:2017poo} between two apparently distinct classes of quantum anomalies. In particular, it was observed that the tensor structures associated with the chiral and conformal anomalies exhibit a double-copy relation. \\
We briefly review the properties of these two anomalies that will be relevant for our analysis.
At the quantum level, anomalies manifest themselves as violations of Ward identities associated with symmetries that are preserved at the classical level. In particular, the conformal anomaly plays a central role in quantum field theory, as it describes the breaking of scale and conformal invariance induced by quantum effects. Beyond its fundamental relevance for the renormalization of quantum field theories, it has found applications in a wide range of contexts, from cosmology and black-hole physics \cite{Mottola:2010gp,Balbinot:1999ri,Hawking:2000bb} to condensed-matter systems \cite{Arouca:2022psl,Chernodub:2021nff}, where it provides a powerful tool for characterizing universal properties of critical phenomena and topological phases.
\\
The conformal anomaly is encoded in the non-vanishing trace of the stress-energy tensor. In four dimensions, in the presence of background gauge and gravitational fields, the most general form of the anomalous trace is \cite{Capper:1974ic,Duff:2020dqb}
\begin{equation}\label{eq:anomtrac}
	g_{\mu \nu}\left\langle T^{\mu \nu}\right\rangle
	=
	b_1 E_4
	+b_2 W^{\mu \nu \rho \sigma} W_{\mu \nu \rho \sigma}
	+b_3 \nabla^2 R
	+b_4 F^{\mu \nu} F_{\mu \nu}
	+f_1 \varepsilon^{\mu \nu \rho \sigma} R_{\alpha \beta \mu \nu} R_{\rho \sigma}^{\alpha \beta}
	+f_2 \varepsilon^{\mu \nu \rho \sigma} F_{\mu \nu} F_{\rho \sigma},
\end{equation}
where the coefficients are determined by the field content of the theory, $W^{\mu \nu \rho \sigma}$ denotes the Weyl tensor and $E_4$ the Euler density.
This anomaly gives rise to an anomalous term in the trace Ward identity of the parity-even stress-tensor three-point function, which takes the form
\begin{equation}\label{eq:anomTTTWI}
	\begin{aligned}
		&\delta_{\mu_3\nu_3}\<T^{\mu_1\nu_1}(\bs{p}_1)T^{\mu_2\nu_2}(\bs{p}_2)T^{\mu_3\nu_3}(\bs{p}_3)\>_{even} \\& \qquad\qquad\qquad =2\<T^{\mu_1\nu_1}(\bs{p}_1)T^{\mu_2\nu_2}(-\bs{p}_1)\>_{even}+2\<T^{\mu_1\nu_1}(\bs{p}_2)T^{\mu_2\nu_2}(-\bs{p}_2)\>_{even}+\mathcal{A}^{\mu_1\nu_1\mu_2\nu_2}
	\end{aligned}
\end{equation}
The anomalous contribution
$\mathcal{A}^{\mu_2\nu_2\mu_3\nu_3}$
is obtained by functionally differentiating the trace anomaly in Eq.~\eqref{eq:anomtrac} twice with respect to the background metric and then Fourier transforming. Here we focus on the Euler contribution, which takes the form\footnote{We factor out the momentum-conservation delta function from the definition, as it has also been removed from the definition of the correlator.}
\begin{equation}
	(2 \pi)^4 \delta^4\left(\bs{p}_1+\bs{p}_2+\bs{p}_3\right) \, \mathcal{A}_{E_4}^{\mu_1\nu_1\mu_2\nu_2}= \int d^4 x_1 \, d^4 x_2\, d^4 x_3\, e^{-i\left(\bs{p}_1 x_1+\bs{p}_2 x_2+\bs{p}_3 x_3\right)} \frac{\delta^2E_4 \left(x_3\right)}{\delta g_{\mu_1 \nu_1}(x_1) \delta g_{\mu_2 \nu_2}(x_2)}
\end{equation}
The full expression is reported in Appendix \ref{appfuncder}.
This is the type-A, topological contribution to the conformal anomaly. As we shall discuss, it is precisely this contribution that is related through a double-copy construction to the chiral anomaly, which is likewise associated with a topological density.
\\
The chiral anomaly represents another fundamental example of a quantum anomaly, associated with the breaking of a classical symmetry by quantum effects.
Since its original discovery in the context of high-energy physics and the pion decay, its implications have extended to a broad range of physical systems, including baryogenesis, the chiral magnetic effect, Weyl and Dirac semimetals and topological phases of matter \cite{Arouca:2022psl,Chernodub:2021nff,Chernodub:2013kya,Landsteiner:2013sja,Frohlich:2023uqc}.
\\
The chiral anomaly is encoded in the non-conservation of an axial current at the quantum level. Its anomalous divergence is given by
\begin{equation}\label{eq:chiranomaly}
	\nabla_\mu \langle  J_A^\mu\rangle=a_1 F\tilde{F}+a_2 R\tilde{R}
	=
	a_1\, \varepsilon^{\mu \nu \rho \sigma} F_{\mu \nu} F_{\rho \sigma}
	+a_2 \, \varepsilon^{\mu \nu \rho \sigma}
	R_{\beta \mu \nu}^{\alpha}
	R_{\alpha \rho \sigma}^{\beta},
\end{equation}
where the anomaly coefficients are again fixed by the matter content of the theory.
The gauge contribution to this anomaly gives rise to the anomalous Ward identity for the parity-odd three-current correlator\footnote{Throughout this paper, we consider for simplicity an abelian current. The extension to the non-abelian case is analogous, with the main modification being the appearance of the corresponding color factor. Note that the two-point functions $\langle JJ\rangle_{{odd}}$ do not need to be included in the Ward identity, since they vanish in $4d$.}
\begin{equation}\label{eq:anomJJJWI}
	\begin{aligned}
		&{p_{3}}_{\mu_3}\<
		J^{\mu_1}(\bs{p}_1)J^{\mu_2}(\bs{p}_2)J^{\mu_3}(\bs{p}_3)\big\rangle_{odd} =a_1 \, \mathcal{A}^{\mu_1\mu_2}_{F\tilde{F}}
	\end{aligned}
\end{equation}
The anomalous tensor structure is obtained by functionally differentiating Eq.~\eqref{eq:chiranomaly} with respect to the background gauge field and then Fourier transforming
\begin{equation}
	(2 \pi)^4 \delta^4\left(\bs{p}_1+\bs{p}_2+\bs{p}_3\right) \, \mathcal{A}_{F\tilde{F}}^{\mu_1\mu_2}= \int d^4 x_1 \, d^4 x_2\, d^4 x_3\, e^{-i\left(\bs{p}_1 x_1+\bs{p}_2 x_2+\bs{p}_3 x_3\right)} \frac{\delta^2{F\tilde{F}} \left(x_3\right)}{\delta A_{\mu_1 }(x_1) \delta A_{\mu_2}(x_2)}
\end{equation}
The Euler and chiral anomaly structures are then related according to
\begin{equation}\label{eq:dcopyanomaly1}
	\mathcal{A}^{\mu_1 \nu_1 \mu_2 \nu_2}_{ E_4}
	\propto
	\delta^{\left(\mu_1\right.}_{\alpha_1} \delta^{\left.\nu_1\right)}_{\beta_1} \delta^{\left(\mu_2\right.}_{\alpha_2} \delta^{\left.\nu_2\right)}_{\beta_2} 
	\mathcal{A}^{\alpha_1 \alpha_2}_{ F \tilde{F}}\mathcal{A}^{\beta_1 \beta_2}_{ F \tilde{F}}
\end{equation}
This double-copy relation, outlined in \cite{Bzowski:2017poo}, originates from the special structure of type-A anomalies in the classification in \cite{Deser:1993yx}. No analogous factorization is found for type-B anomalies, such as the Weyl-squared contribution to the trace anomaly.
\\
Equation~\eqref{eq:dcopyanomaly1} relates the anomalous trace sector of the $\langle TTT\rangle_{even}$ correlator to the anomalous longitudinal sector of the $\langle JJJ\rangle_{odd}$ correlator. In the following sections, we will show that related double-copy structures also emerge in the transverse-traceless sectors of the corresponding correlators.
\\
As we can notice, the left-hand side of Eq.~\eqref{eq:dcopyanomaly1} is parity even, consistently with the fact that the right-hand side is the product of two parity-odd structures.
A natural question is whether a similar construction can be formulated for parity-odd gravitational structures. In the present work, we show that this is indeed the case and derive a new mixed-parity double-copy relation. To this end, in addition to the parity-odd structure
$\mathcal{A}_{F\widetilde{F}}$, we introduce the parity-even gauge-field structure
\begin{equation}
	(2 \pi)^4 \delta^4\left(\bs{p}_1+\bs{p}_2+\bs{p}_3\right) \, \mathcal{A}_{F^2}^{\mu_1\mu_2}=\int d^4 x_1 \, d^4 x_2\, d^4 x_3\,e^{-i\left(\bs{p}_1 x_1+\bs{p}_2 x_2+\bs{p}_3 x_3\right)} \frac{\delta^2{F^2} \left(x_3\right)}{\delta A_{\mu_1 }(x_1) \delta A_{\mu_2}(x_2)}
\end{equation}
together with the parity-odd gravitational structure generated by the Pontryagin density
\begin{equation}
	(2 \pi)^4 \delta^4\left(\bs{p}_1+\bs{p}_2+\bs{p}_3\right) \, \mathcal{A}_{R\tilde{R}}^{\mu_1\nu_1\mu_2\nu_2}= \int d^4 x_1 \, d^4 x_2\, d^4 x_3\, e^{-i\left(\bs{p}_1 x_1+\bs{p}_2 x_2+\bs{p}_3 x_3\right)} \frac{\delta^2{R\tilde{R}} \left(x_3\right)}{\delta g_{\mu_1\nu_1 }(x_1) \delta g_{\mu_2\nu_2}(x_2)}
\end{equation}
Their full expressions are reported in Appendix \ref{appfuncder}.
We then find that these structures obey the mixed-parity double-copy relation
\begin{equation}\label{eq:dcopyanomaly2}
	\mathcal{A}^{\mu_1 \nu_1 \mu_2 \nu_2}_{R \tilde{R} }
	\propto
	\delta^{\left(\mu_1\right.}_{\alpha_1} \delta^{\left.\nu_1\right)}_{\beta_1} \delta^{\left(\mu_2\right.}_{\alpha_2} \delta^{\left.\nu_2\right)}_{\beta_2} 
	\mathcal{A}^{\alpha_1 \alpha_2}_{ F \tilde{F}}\mathcal{A}^{\beta_1 \beta_2}_{ F^2}
\end{equation}
In this case, the product of one parity-odd and one parity-even gauge structure reproduces the parity-odd gravitational structure.
\\
Equations~\eqref{eq:dcopyanomaly1} and \eqref{eq:dcopyanomaly2} may therefore be interpreted as double-copy relations connecting gauge and gravitational anomaly densities. However, the same identities admit a complementary interpretation in terms of three-point vertices involving a scalar or pseudoscalar field coupled to gauge and gravitational densities.
\\
In particular, Eq.~\eqref{eq:dcopyanomaly1} can be rewritten as
\begin{equation} \label{eq:dcopyscalarA1}
	\mathcal{A}^{\mu_1 \nu_1 \mu_2 \nu_2}_{\phi E_4}
	\propto 
	\delta^{\left(\mu_1\right.}_{\alpha_1} \delta^{\left.\nu_1\right)}_{\beta_1} \delta^{\left(\mu_2\right.}_{\alpha_2} \delta^{\left.\nu_2\right)}_{\beta_2} 
	\mathcal{A}^{\alpha_1 \alpha_2}_{\phi F \tilde{F}}\mathcal{A}^{\beta_1 \beta_2}_{\phi F \tilde{F}}
\end{equation} 
The left-hand side describes a three-point interaction between a scalar field and two gravitons generated by a coupling of the form $\phi E_4$
\begin{equation}\label{eq:defaphie4}
	(2 \pi)^4 \delta^4\left(\bs{p}_1+\bs{p}_2+\bs{p}_3\right) \, \mathcal{A}_{\phi E_4}^{\mu_1\nu_1\mu_2\nu_2}=\int d^4 x_1 \, d^4 x_2\, d^4 x_3\, e^{-i\left(\bs{p}_1 x_1+\bs{p}_2 x_2+\bs{p}_3 x_3\right)} \frac{\delta^3\mathcal{S}_{\phi E_4} }{\delta g_{\mu_1 \nu_1}(x_1) \delta g_{\mu_2 \nu_2}(x_2)\delta \phi(x_3)}
\end{equation}
where we have introduced the interaction action
\begin{equation}
	\mathcal{S}_{\phi E_4}\equiv \int d^4x\,  \phi \, E_4 
\end{equation}
The two factors on the right-hand side of Eq.~\eqref{eq:dcopyscalarA1} instead correspond to three-point interactions between a pseudoscalar field and two gauge fields, generated by a coupling of the form $\phi F\widetilde{F}$. These interactions are analogous to the familiar axion--photon coupling and can be defined in a manner analogous to Eq.~\eqref{eq:defaphie4}.
\\
Similarly, the mixed-parity relation becomes
\begin{equation} \label{eq:dcopyscalarA2}
	\mathcal{A}^{\mu_1 \nu_1 \mu_2 \nu_2}_{\phi R \tilde{R} }
	\propto
	\delta^{\left(\mu_1\right.}_{\alpha_1} \delta^{\left.\nu_1\right)}_{\beta_1} \delta^{\left(\mu_2\right.}_{\alpha_2} \delta^{\left.\nu_2\right)}_{\beta_2} 
	\mathcal{A}^{\alpha_1 \alpha_2}_{ \phi F \tilde{F}}\mathcal{A}^{\beta_1 \beta_2}_{ \phi F^2}
\end{equation}
Here the parity-odd gravitational interaction generated by
$\phi R\widetilde{R}$
is obtained as the double copy of the parity-odd gauge interaction
$\phi F\widetilde{F}$
and the parity-even interaction
$\phi F^2$.
\\
As we will show in detail in the following section, Eqs.~\eqref{eq:dcopyscalarA1} and \eqref{eq:dcopyscalarA2} are the three-point manifestations of more general double-copy relations between conformal correlators involving scalar or pseudoscalar operators.

\section{Double copy relations for the $\langle JJO\rangle$ and $\langle TTO\rangle$ correlators}\label{sec:dcopyttojjo}
Before turning to the main subject of this paper, namely the chiral $\langle JJJ\rangle_{odd}$  and conformal $\langle TTT\rangle_{even}$ interactions, it is instructive to consider a simpler class of examples involving three-point correlators with a marginal scalar or pseudoscalar operator.
In general, the double-copy construction is formulated by contracting the conformal correlators with polarisation vectors and uplifting the kinematics to $d+1$ dimensions, following the procedure of \cite{Farrow:2018yni,Lipstein:2019mpu}, which provides a map between conformal correlators and scattering amplitudes. We will outline this construction in the next section. Here, however, this additional step is not necessary, since the double-copy structure can already be identified directly at the level of the correlators themselves.
\\
In this section, we investigate the following double-copy relation\footnote{In this equation, we use a schematic notation. Since the correlators can be parametrized by multiple independent constants, verifying the double-copy relations requires selecting specific sectors of the correlators. The same consideration applies to the relation between $\langle TTT\rangle$ and $\langle JJJ\rangle$, for which a suitable kinematic limit is also required, as we will see in later sections.}
\begin{equation}
	\langle TTO_4\rangle\propto \langle JJO_4\rangle\langle JJO_4\rangle
\end{equation}
where $O_4$ is a scalar operator of conformal dimension $\Delta=d=4$.
Decomposing the correlators into their parity-even and parity-odd components
\begin{equation}
	\langle TTO_4\rangle=\langle TTO_4\rangle_{even} +\langle TTO_4\rangle_{odd} \qquad\qquad \langle JJO_4\rangle=\langle JJO_4\rangle_{even}+\langle JJO_4\rangle_{odd},
\end{equation}
one obtains
\begin{equation}\label{eq:dcopyscalari}
	\begin{aligned}
		& &&\langle TTO_4\rangle_{even} \propto \langle JJO_4\rangle_{even} \langle JJO_4\rangle_{even},\\
		&\qquad && \langle TTO_4\rangle_{even} \propto  \langle JJO_4\rangle_{odd} \langle JJO_4\rangle_{odd} ,\\
		& \qquad && \langle TTO_4\rangle_{odd} \propto \langle JJO_4\rangle_{even} \langle JJO_4\rangle_{odd} 
	\end{aligned}
\end{equation}
The first relation was established in \cite{Farrow:2018yni,Lipstein:2019mpu}. In what follows, we therefore concentrate on the second and third relations.
We begin by reviewing the relevant correlators.
\\
The parity-odd sector of the correlators was derived in \cite{Coriano:2024ssu} by solving the conformal Ward identities\footnote{In \cite{Coriano:2024ssu}, no renormalization of the correlators was performed. One might therefore expect these results to correspond to contact terms. A consistent renormalization of these conformal correlators and the associated 3K integrals remains an open problem, as it requires a careful treatment of the Levi-Civita tensor, which is intrinsically defined in four dimensions.}. 
We will not detail the derivation of the solution presented below, as the underlying procedure will be discussed more extensively in the following sections, where we apply it to correlators such as the $\langle JJJ\rangle_{odd}$.
In $d=4$, the results are
\begin{equation}
	\begin{aligned}
		&\langle JJO_4\rangle_{odd} = c_1 \varepsilon^{\mu_1 \mu_2 \bs{p}_1 \bs{p}_2}=\mathcal{A}^{\mu_1 \mu_2}_{\phi F \tilde{F}}
	\end{aligned}
\end{equation}
and
\begin{equation}
	\begin{aligned}
		& \left\langle T^{\mu_1 \nu_1} T^{\mu_2 \nu_2}O_{4}\right\rangle_{\text {odd }}= \\
		 & \qquad\qquad c_1\left[\varepsilon^{\nu_1 \nu_2 \bs{p}_1 \bs{p}_2}\left(\left(\bs{p}_1 \cdot \bs{p}_2\right) \delta^{\mu_1 \mu_2}-\bs{p}_1^{\mu_2} \bs{p}_2^{\mu_1}\right)+\left(\mu_1 \leftrightarrow \nu_1\right)+\left(\mu_2 \leftrightarrow \nu_2\right)+\binom{\mu_1 \leftrightarrow \nu_1}{\mu_2 \leftrightarrow \nu_2}\right] =\mathcal{A}^{\mu_1 \nu_1 \mu_2 \nu_2}_{\phi R \tilde{R} }
	\end{aligned}
\end{equation}
As discussed in \cite{Coriano:2024ssu,Coriano:2023cvf}, two physically relevant realizations of the operator $O_4$ are given by $\nabla \cdot J_A$ or $T^\mu_\mu$, corresponding to the chiral and conformal anomalies, respectively.
\\
The parity-even sector, on the other hand, was analyzed in \cite{Bzowski:2018fql}. The general solution of the correlators is parametrized by several independent coefficients. In particular, after renormalization, the correlators can contain local contributions arising from finite counterterms, whose coefficients are fixed by the choice of renormalization scheme. These are precisely the terms relevant for our discussion and are highlighted below
\begin{equation}\label{eq:jjoeven}
	\langle J^{\mu_1}J^{\mu_2}O_4\rangle_{even} =  \mathcal{A}^{\mu_1\mu_2}_{\phi F^2}+\dots
\end{equation}
\begin{equation}\label{eq:ttoeven}
	\langle T^{\mu_1\nu_1}T^{\mu_2\nu_2}O_4\rangle_{even} =  \mathcal{A}^{\mu_1\nu_1 \mu_2\nu_2}_{\phi E_4}+\dots
\end{equation}
See Appendix \ref{appJJOTTO} for the complete expressions.
Here, the ellipses denote additional conformally invariant contributions that are not relevant for our discussion. Since their coefficients are arbitrary, we can consistently restrict our attention to the sector in which they vanish. Under this restriction, the resulting conformal correlators satisfy the second and third relations in Eq.~\eqref{eq:dcopyscalari}, corresponding respectively to Eqs.~\eqref{eq:dcopyscalarA1} and \eqref{eq:dcopyscalarA2}.

\section{Review: from CFT correlators to flat-space amplitudes}  \label{review}
In this section, we review some basic results on momentum-space conformal correlators in $d$ Euclidean dimensions \cite{Bzowski:2013sza, Bzowski:2017poo, Bzowski:2018fql} and their relation to scattering amplitudes in $(d+1)$-dimensional Minkowski space. For parity-even correlators, this correspondence was established in \cite{Farrow:2018yni} for odd dimensions and in \cite{Lipstein:2019mpu} for even dimensions. The parity-odd case was investigated in \cite{Jain:2021qcl}, specifically in three dimensions, where no anomalies are present.
\\
A general correlator can be decomposed into trace, longitudinal and transverse-traceless components. The trace and longitudinal parts can be reconstructed from lower-point functions using the corresponding (anomalous) Ward identities. The transverse-traceless sector, on the other hand, can be expressed as a sum of tensor structures, each multiplied by a scalar form factor. For three-point correlators, these form factors depend only on the magnitudes of the external momenta
\begin{equation}
	p_i=+\sqrt{\bs{p}_i^2}, \qquad i \in \left\{ 1,2,3\right\},
\end{equation}
since momentum conservation allows scalar products such as $\bs{p}_1\cdot\bs{p}_2$ to be expressed as $(p_3^2-p_1^2-p_2^2)/2$, {\it etc.}
For physical kinematics, these momentum magnitudes satisfy the triangle inequalities $0 \leq p_i \leq p_j+p_k$. In our computations, it will be useful to introduce the following notation for combinations of the external momenta
\begin{equation}
			c_{123}\equiv p_1p_2p_3, \qquad \qquad J^2\equiv \left(p_1+p_2+p_3\right)\left(-p_1+p_2+p_3\right)\left(p_1-p_2+p_3\right)\left(p_1+p_2-p_3\right)
\end{equation}
Since our interest lies in scattering amplitudes, we need to contract all tensor indices of the correlators with transverse polarisation vectors $\bs{\epsilon}_i=\bs{\epsilon}(\bs{p}_i)$ satisfying
\begin{equation}
\bs{\epsilon}_i\cdot \bs{p}_i = 0, \qquad
\bs{\epsilon}_i\cdot\bs{\epsilon}_i = 0.
\end{equation}
Note that for gravitons, we can write polarisation tensors in terms of polarisation vectors as $\bs{\epsilon}^{\mu \nu}_i=\bs{\epsilon}^\mu_i \bs{\epsilon}^\nu_i$. 
To establish the connection with scattering amplitudes, we lift the kinematics to $(d+1)$-dimensional Minkowski space by introducing the bulk null momenta and polarisation vectors
\begin{equation}\label{eq:fourtofivemomentapolariz}
	p^\mu_i = (p_i,\bs{p}_i), \qquad \epsilon_i^\mu = (0,\bs{\epsilon}_i).
\end{equation}
Scalar products involving the spatial polarisation vectors are then promoted to their bulk counterparts through the replacements $\bs{\epsilon}_i\cdot \bs{p}_j \rightarrow \epsilon_i\cdot p_j$ and $\bs{\epsilon}_i\cdot\bs{\epsilon}_j\rightarrow \epsilon_i\cdot\epsilon_j$.
One should note that, unlike the original $d$-dimensional momenta, the bulk momenta do not, in general, obey momentum conservation. Instead, their sum is given by
\begin{equation}
\sum_{i=1}^3 p_i^\mu = (E, \bs{0}),
\end{equation}
where the total bulk energy is
\begin{equation}\label{Edef}
E=p_1+p_2+p_3.
\end{equation}
Therefore, we will consider the leading behaviour of CFT correlators in the limit $E\rightarrow 0$ for which energy conservation is restored. 
This limit is naturally interpreted as the flat-space limit, either of $(d+1)$-dimensional anti-de Sitter space \cite{Gary:2009ae, Penedones:2010ue, Raju:2012zr,Marotta:2024sce} or, alternatively, of $(d+1)$-dimensional de Sitter space \cite{Maldacena:2011nz, Arkani-Hamed:2015bza, Farrow:2018yni,Lipstein:2019mpu, Arkani-Hamed:2018kmz}.
\\
In this limit, the coefficients of the leading singularities of $d$-dimensional CFT correlators are precisely the $(d+1)$-dimensional flat-space scattering amplitudes, which exhibit a double-copy structure. In \cite{Farrow:2018yni,Lipstein:2019mpu}, it was shown that, in the flat-space limit, three-point correlators of stress tensors and conserved currents reduce to linear combinations of the following gauge-theory and gravitational amplitudes
\begin{equation}
	\mathcal{A}_{EG}=(\mathcal{A}_{YM})^{2},\qquad \mathcal{A}_{\phi R^{2}}^{222}=\mathcal{A}_{F^{3}}\mathcal{A}_{YM},\qquad \mathcal{A}_{W^{3}}=(\mathcal{A}_{F^{3}})^{2}
	\label{gravityamp}
\end{equation}
which are related by the double copy. 
In particular, $\mathcal{A}_{EG}$ denotes the three-graviton amplitude in Einstein gravity, $\mathcal{A}_{W^{3}}$ is the corresponding amplitude in Weyl-cubed gravity, and $\mathcal{A}_{\phi R^{2}}^{222}$ is the three-graviton amplitude (the superscript $222$ indicates three external gravitons) in the curvature-squared theory of gravity coupled to scalars constructed in \cite{Broedel:2012rc} (see also \cite{Johansson:2017srf}). The corresponding gauge-theory amplitudes are
\begin{equation}
	\mathcal{A}_{YM}=\epsilon_{1}\cdot\epsilon_{2}\,\,\epsilon_{3}\cdot p_{1}+\mathrm{cyclic},\qquad \mathcal{A}_{F^{3}}=\epsilon_{1}\cdot p_{2}\,\,\epsilon_{2}\cdot p_{3}\,\,\epsilon_{3}\cdot p_{1},
	\label{gaugeamp}
\end{equation}
where $\mathcal{A}_{YM}$ is the three-gluon Yang--Mills amplitude and $\mathcal{A}_{F^{3}}$ is the corresponding amplitude in a higher-derivative gauge theory containing an $F^3$ interaction \cite{Johansson:2017srf}.\\
In this paper, we focus specifically on four-dimensional CFT correlators. Accordingly, the flat-space amplitudes relevant to our discussion are five-dimensional objects. We therefore stress that the quantities appearing in Eqs.~\eqref{gravityamp} and~\eqref{gaugeamp} should not be confused with the four-dimensional quantities introduced previously in Sections~\ref{sec:dcopyanomaly} and~\ref{sec:dcopyttojjo}. From this point onwards, we use $\mathcal{A}$ exclusively to denote five-dimensional amplitudes, so that no ambiguity should arise.
In addition to the parity-even amplitudes above, we will also consider the following parity-odd gauge-theory amplitude
\begin{equation} 
	\mathcal{A}_{AF\tilde{F}}=\varepsilon^{p_1 p_2 \epsilon_1\epsilon_2\epsilon_3 }
\end{equation}
which arises from the cubic Chern--Simons interaction. 

\subsection{3K integrals and their limits}

As discussed in the previous section, the transverse-traceless sector of correlators is the one relevant to our purposes, as it is the component that survives upon contraction with polarization vectors. This sector
can be decomposed into a sum of tensor structures, each multiplied by a scalar form factor.
In general, the form factors are arbitrary functions of the momenta; however, for two- and three-point functions, they can be determined completely by imposing conformal invariance. In momentum space, the general solution to the conformal Ward identities for three-point correlators can be expressed as specific linear combinations of 3K integrals
\cite{Bzowski:2013sza,Bzowski:2015pba,Bzowski:2015yxv}
 \begin{align}
 	I_{\alpha \{ \beta_1, \beta_2, \beta_3 \}}(p_1, p_2, p_3) & = \int_0^\infty \mathrm{d} x \: x^\alpha \prod_{i=1}^3 p_i^{\beta_i} K_{\beta_i}(p_i x), \label{tripleK} 
 \end{align}
 where $K_{\beta_i}$ is a modified Bessel function of the second kind
 \begin{equation}
 	K_\beta(x)=\frac{\pi}{2} \frac{I_{-\beta}(x)-I_\beta(x)}{\sin (\nu \pi)}, \qquad \nu \notin \mathbb{Z} \qquad\qquad I_\beta(x)=\left(\frac{x}{2}\right)^\beta \sum_{k=0}^{\infty} \frac{1}{\Gamma(k+1) \Gamma(\nu+1+k)}\left(\frac{x}{2}\right)^{2 k}
 \end{equation}
The arguments of the 3K integral are magnitudes of momenta $p_i$, with $i=1, 2, 3$. In addition, 3K integrals depend on four parameters: the power $\alpha$ of the integration variable $x$, and the three Bessel function indices $\beta_i$. 
As we will see, the various integrals that arise in our analysis, corresponding to different values of $\alpha$ and $\beta_i$, can be constructed by applying differential operators to the master integral $I_{1{000}}$,  see Appendix~\ref{app3kexpression} and \cite{Bzowski:2015yxv}. For physical momentum configurations satisfying the triangle inequalities $p_i \leq p_j+p_k$, working in $d=4$,  such master integral is equivalent to a one-loop triangle 
\begin{equation}
	I_{1\{000\}}\left(p_1, p_2, p_3\right)=\frac{1}{4 \pi^2} \int \frac{\mathrm{~d}^4 \ell}{\ell^2\left(\ell+p_1\right)^2\left(\ell-p_3\right)^2}
\end{equation}
An important subtlety in the evaluation of 3K integrals is that they can develop divergences for specific values of their indices. In general, a 3K integral diverges whenever its indices satisfy
\begin{equation}
	\alpha+1 \pm \beta_1\pm \beta_2 \pm \beta_3 = -2n
\end{equation}
for any independent choice of the $\pm$ signs and any non-negative integer $n$. To regulate these divergences, one introduces infinitesimal shifts in the operator and spacetime dimensions, and hence in the indices $\alpha$ and ${\beta_i}$ parametrising the 3K integrals. The divergent terms can then be isolated and removed by adding covariant local counterterms. Finally, the regulator is removed to obtain the renormalised correlators. This procedure is required, in particular, for the conformal-anomaly correlator $\langle TTT\rangle_{\mathrm{even}}$, which we discuss in a later section.
\\
We now turn to the flat-space limit of 3K integrals.
For CFTs in odd spacetime dimensions, the relevant 3K integrals involve half-integer indices, and the resulting form factors reduce to simple rational functions of the momentum magnitudes. The flat-space limit is therefore obtained simply by extracting the leading behaviour as $E\rightarrow0$.
The situation is more subtle in even spacetime dimensions. First, the form factors diverge and therefore require regularisation and renormalisation. Second, the renormalised form factors develop a richer analytic structure involving branch cuts. Consequently, the analytic continuation required to define the flat-space limit must be specified with care.
\\
The flat-space limit of 3K integrals has been studied in both odd and even dimensions in \cite{Farrow:2018yni} and \cite{Lipstein:2019mpu}, respectively. Here, we present an asymptotic analysis relevant to both the odd- and even-dimensional case, referring the reader to the latter reference for details of the derivation.
\\
The key step is to analytically continue one of the momentum magnitudes according to
\begin{equation}
	p_3=\left|p_3\right| e^{i \theta}, \quad 0 \leq \theta \leq \pi,
\end{equation}
After continuing from
$\theta=0$ to $\theta=\pi$, the flat-space limit then corresponds to sending
\begin{equation}
	\left|p_3\right| \rightarrow p_1+p_2 .
\end{equation}
Using the analytic continuation formula
\begin{equation}
	K_\nu\left(e^{i \pi} x\right)=e^{-i \pi \nu} K_\nu(x)-i \pi I_\nu(x), \quad x \in \mathbb{R}^{+}, \quad \nu \in \mathbb{Z},
\end{equation}
we find
\begin{equation}
	\begin{aligned}
		I_{\alpha\{\beta_{1}\beta_{2}\beta_{3}\}}(p_{1},p_{2},p_{3})\Big|_{\theta=\pi}&=I_{\alpha\{\beta_{1}\beta_{2}\beta_{3}\}}(p_{1},p_{2},|p_{3}|)\nn\\&\quad
		-i \pi p_{1}^{\beta_{1}}p_{2}^{\beta_{2}}p_{3}^{\beta_{3}}\int_{0}^{\infty}\mathrm{d} x\,x^{\alpha}K_{\beta_{1}}(p_{1}x)K_{\beta_{2}}(p_{2}x)I_{\beta_{3}}(|p_{3}|x),
		\label{tripleKcont}
	\end{aligned}
\end{equation}
where the phase $e^{-i\pi\beta_3}$ arising from the continuation of the Bessel function is cancelled by the corresponding phase from the continuation of $p_3^{\beta_3}$ in the definition \eqref{tripleK}.
We can now extract the flat-space limit by examining the asymptotic behaviour of the integrand in the analytically continued expression. Physically, this limit corresponds to probing the deep interior of the bulk, where
$p_i x \gg 1$. In this regime, the Bessel functions can be replaced by their asymptotic behaviours
\begin{equation}
K_{\beta}(p_i x)\rightarrow\sqrt{\frac{\pi}{2p_i x}}\,e^{-p_ix},\qquad I_{\beta}(p_i x)\rightarrow\sqrt{\frac{1}{2\pi p_i x}}\,e^{p_i x},
\end{equation}
which leads to
\begin{equation}
	\lim_{E\rightarrow 0} I_{\alpha\{\beta_{1}\beta_{2}\beta_{3}\}}(p_{1},p_{2},p_{3})\Big|_{\theta=\pi}=
	-\frac{\pi^{3/2}\Gamma(\alpha-1/2)\Pi_{i=1}^{3}p_{i}^{\beta_{i}-1/2}}{\sqrt{8}E^{\alpha-1/2}}.
	\label{flat}
\end{equation}
The leading singular behaviour arises entirely from the $KKI$ integral in \eqref{tripleKcont}, since the 3K integral itself remains finite for collinear momentum configurations \cite{Bzowski:2013sza}.

\section{The chiral-anomaly correlator}\label{sec:chiranomalcorrel}
In this section, we investigate the chiral-anomaly correlator and show how it is related to a five-dimensional flat-space amplitude.
We start by recalling Eq.~\eqref{eq:chiranomaly}. The gauge field strength $F_{\mu\nu}$ may correspond to either a vector or an axial gauge field. In principle, one can consider a theory in which both fields are present, so that both contributions appear in the chiral anomaly equation. Restricting to flat spacetime, one can therefore write
\begin{equation}
	\partial_\mu \langle  J_A^\mu\rangle =\epsilon^{\mu \nu \rho \sigma}\left[a_1 F^V_{ \mu \nu} F^V_ {\rho \sigma}+a'_1 F^A_{ \mu \nu} F^A_{ \rho \sigma}\right]
\end{equation}
Here, $F^V_{\mu\nu}$ and $F^A_{\mu\nu}$ denote the field strengths of the vector gauge field $V_\mu$ and the axial gauge field $A_\mu$, respectively.
For a fermion coupled with equal charge to both the vector and axial gauge fields, the corresponding anomaly coefficients satisfy
\begin{equation}
	\label{eq:idanomcoef}
	a'_1=\frac{a_1}{3}
\end{equation}
At the perturbative level, the correlators containing one and three axial currents are related by
\begin{equation}\label{eq:distribuzanomal}
	\langle  J_AJ_AJ_A\rangle=\frac{1}{3}\bigg[ \langle  J_VJ_VJ_A\rangle+ \langle  J_VJ_AJ_V\rangle+\langle  J_AJ_VJ_V\rangle \bigg]
\end{equation}
which expresses the three-axial-current correlator as the average over the three possible orderings of the correlator containing one axial and two vector currents.
\\
In \cite{Coriano:2023hts}, both the correlator containing one axial current and the correlator containing three axial currents were reconstructed by solving their anomalous and conformal Ward identities. As an illustrative example, let us consider the $\langle J_VJ_VJ_A\rangle$ correlator. Its reconstruction begins by imposing the conservation of the vector currents together with the anomalous Ward identity for the axial current
\begin{equation}
	\begin{aligned}
		& \bs{p}_{i \mu_i}\left\langle J_V^{\mu_1}\left(\bs{p}_1\right) J_V^{\mu_2}\left(\bs{p}_2\right) J_A^{\mu_3}\left(\bs{p}_3\right)\right\rangle=0, \quad i=1,2 \\
		& \bs{p}_{3 \mu_3}\left\langle J_V^{\mu_1}\left(\bs{p}_1\right) J_V^{\mu_2}\left(\bs{p}_2\right) J_A^{\mu_3}\left(\bs{p}_3\right)\right\rangle=-8 \, a_1 i\, \varepsilon^{\bs{p}_1 \bs{p}_2 \mu_1 \mu_2}
	\end{aligned}
\end{equation}
The most general solution to these Ward identities can be decomposed into a longitudinal contribution, which is completely fixed by the anomaly, and a transverse contribution, which is left unconstrained
\begin{equation}
	\begin{aligned}
		&\left\langle J_V^{\mu_1}\left(\bs{p}_1\right) J_V^{\mu_2}\left(\bs{p}_2\right) J_A^{\mu_3}\left(\bs{p}_3\right)\right\rangle= -8 i a_1 \frac{\bs{p}_{3}^{\mu_3}}{p_3^2} \varepsilon^{\bs{p}_1 \bs{p}_2 \mu_1 \mu_2}+\pi_{\alpha_1}^{\mu_1}\left(\bs{p}_1\right) \pi_{\alpha_2}^{\mu_2}\left(\bs{p}_2\right) \pi_{\alpha_3}^{\mu_3}\left(\bs{p}_3\right) X^{\alpha_1 \alpha_2 \alpha_3}	
		\end{aligned}
\end{equation}
Here, the transverse part is parametrized by an undetermined rank-three tensor $X^{\alpha_1 \alpha_2 \alpha_3}$ and 
\begin{equation}
	\pi_\alpha^\mu(\bs{p})=\delta_\alpha^\mu-\frac{\bs{p}^\mu \bs{p}_\alpha}{\bs{p}^2}
\end{equation}
denotes the transverse projector.
The tensor $X^{\alpha_1\alpha_2\alpha_3}$ admits a decomposition in terms of two independent tensor structures multiplied by form factors $A_i$
\begin{equation}\label{eq:decomtransvavv}
	\begin{aligned}
 X^{\alpha_1 \alpha_2 \alpha_3}=A_1\left(p_1, p_2, p_3\right) \varepsilon^{\bs{p}_1 \bs{p}_2 \alpha_1 \alpha_2} \bs{p}_1^{\alpha_3} +A_2\left(p_1, p_2, p_3\right) \varepsilon^{\bs{p}_1 \alpha_1 \alpha_2 \alpha_3}-A_2\left(p_2, p_1, p_3\right) \varepsilon^{\bs{p}_2 \alpha_1 \alpha_2 \alpha_3}
	\end{aligned}
\end{equation}
where the Bose symmetry under the exchange of the two vector currents requires $A_1(p_1,p_2,p_3) =-A_1(p_2,p_1,p_3)$. For the $\langle J_AJ_AJ_A\rangle$ correlator, the enhanced Bose symmetry under permutations of all three currents imposes an additional constraint, uniquely determining $A_1$ in terms of $A_2$.

\subsection{The conformal solutions and the flat-space limit}
The form factors $A_i$ introduced in Eq.~\eqref{eq:decomtransvavv} encode the transverse part of the correlator and are, a priori, arbitrary functions of the external momenta. Conformal symmetry provides additional constraints that determine these functions through the dilatation Ward identity
\begin{equation}
	\left(\sum_{i=1}^3 \Delta_i-2 d-\sum_{i=1}^2 \bs{p}_i^\mu \frac{\partial}{\partial \bs{p}_i^\mu}\right)\left\langle J_V^{\mu_1}\left(\bs{p}_1\right) J_V^{\mu_2}\left(\bs{p}_2\right) J_A^{\mu_3}\left(\bs{p}_3\right)\right\rangle=0 .
\end{equation}
and the special conformal Ward identity
\begin{equation}
	\begin{aligned}
		0=  \sum_{j=1}^2\left[-2 \frac{\partial}{\partial \bs{p}_{j \kappa}}-2 \bs{p}_j^\alpha \frac{\partial^2}{\partial \bs{p}_j^\alpha \partial \bs{p}_{j \kappa}}+\bs{p}_j^\kappa \frac{\partial^2}{\partial \bs{p}_j^\alpha \partial \bs{p}_{j \alpha}}\right]\left\langle J_V^{\mu_1}\left(\bs{p}_1\right) J_V^{\mu_2}\left(\bs{p}_2\right) J_A^{\mu_3}\left(\bs{p}_3\right)\right\rangle &\\
		 +2\left(\delta^{\mu_1 \kappa} \frac{\partial}{\partial \bs{p}_1^{\alpha_1}}-\delta_{\alpha_1}^\kappa \frac{\partial}{\partial \bs{p}_{1 \mu_1}}\right)\left\langle J_V^{\alpha_1}\left(\bs{p}_1\right) J_V^{\mu_2}\left(\bs{p}_2\right) J_A^{\mu_3}\left(\bs{p}_3\right)\right\rangle &\\
		+2\left(\delta^{\mu_2 \kappa} \frac{\partial}{\partial \bs{p}_2^{\alpha_2}}-\delta_{\alpha_2}^\kappa \frac{\partial}{\partial \bs{p}_{2 \mu_2}}\right)\left\langle J_V^{\mu_1}\left(\bs{p}_1\right) J_V^{\alpha_2}\left(\bs{p}_2\right) J_A^{\mu_3}\left(\bs{p}_3\right)\right\rangle&.
	\end{aligned}
\end{equation}
The solution to this system of equations is unique and is completely determined by the anomaly coefficients $a_1$ and $a'_1$. In particular, after contracting the correlator with the polarisation vectors, the result can be expressed in terms of 3K integrals as \cite{Coriano:2023hts}
\begin{equation}\label{eq:avvconf}
	\left\langle J_V\left(\bs{p}_1\right) J_V\left(\bs{p}_2\right) J_A\left(\bs{p}_3\right)\right\rangle=8 i a_1 \bigg[{p}_2^2 I_{3\{101\}} \varepsilon^{\bs{p}_1 \bs{\epsilon}_1 \bs{\epsilon}_2 \bs{\epsilon}_3}-{p}_1^2 I_{3\{011\}} \varepsilon^{\bs{p}_2 \bs{\epsilon}_1 \bs{\epsilon}_2 \bs{\epsilon}_3}\bigg]
\end{equation}
and 
\begin{equation}\label{eq:aaaconf}
	\begin{aligned}
		&\left\langle J_A\left(\bs{p}_1\right) J_A\left(\bs{p}_2\right) J_A\left(\bs{p}_3\right)\right\rangle =  
		24 i a'_1 \bigg[\left(I_{3\{101\}} {p}_2^2-\frac{1}{3} \right) \varepsilon^{\bs{p}_1 \bs{\epsilon}_1 \bs{\epsilon}_2 \bs{\epsilon}_3}-\left(I_{3\{011\}} {p}_1^2-\frac{1}{3}\right) \varepsilon^{\bs{p}_2 \bs{\epsilon}_1 \bs{\epsilon}_2 \bs{\epsilon}_3}\bigg]
	\end{aligned}
\end{equation}
The 3K integrals appearing in these expressions can be written in terms of derivatives of the master integral $I_{1\{000\}}$, as detailed in Appendix~\ref{app3kexpression}.
It is worth emphasizing that these solutions are entirely generated by the anomaly: in the absence of the anomaly, namely in the limit $a_1\rightarrow0$ and $a'_1\rightarrow0$, both correlators vanish.
Moreover, using the properties of the 3K integrals, one can verify that these solutions satisfy Eq$.$ \eqref{eq:distribuzanomal}, once the relation between the anomaly coefficients in Eq.~\eqref{eq:idanomcoef} is imposed. 
\\
The next step is to reinterpret the four-dimensional conformal correlators as five-dimensional flat-space amplitudes. Following the procedure outlined in Section \ref{review}, we take the flat-space limit of the correlators. Remarkably, the correlators containing one and three axial currents reduce to the same five-dimensional amplitude. We therefore denote both of these correlators by $\langle JJJ\rangle_{{odd}}$. More precisely, using Eq.~\eqref{flat}, we find
\begin{equation}\label{eq:avvlimitflat}
	\lim _{E \rightarrow 0}\frac{E^{{5}/{2}}}{c_{123}^{{1}/{2}}}\, \langle J J J\rangle_{odd}  \propto- \, {p}_2\,  \varepsilon^{\bs{p}_1\bs{\epsilon}_1\bs{\epsilon}_2\bs{\epsilon}_3 }+ {p}_1 \,\varepsilon^{\bs{p}_2 \bs{\epsilon}_1\bs{\epsilon}_2\bs{\epsilon}_3} =\varepsilon^{p_1 p_2 \epsilon_1\epsilon_2\epsilon_3 }= \mathcal{A}_{AF\tilde{F}}
\end{equation}
Note that we have switched from four-dimensional to five-dimensional notation, introducing the five-dimensional Levi-Civita tensor. The equality follows from the fact that the time component of the five-dimensional Levi-Civita tensor can only be saturated by the momenta, since the polarisation vectors have vanishing time components, see Eq.~\eqref{eq:fourtofivemomentapolariz}. \\
The resulting expression can be identified with the five-dimensional amplitude $\mathcal{A}_{AF\tilde{F}}$, associated with the cubic Chern--Simons action
\begin{equation}
	\mathcal{S}_{CS}=\int d^5x \, \varepsilon^{\mu_1\mu_2\mu_3\mu_4\mu_5}A_{\mu_1}F_{\mu_2\mu_3}F_{\mu_4\mu_5}
\end{equation}
Therefore, the flat-space limit provides a direct map between the four-dimensional chiral-anomaly correlators in CFT and a five-dimensional Chern--Simons interaction.\footnote{Such Chern--Simons contributions to anomalous current correlators are familiar from holography \cite{Witten:1998qj,Freedman:1998tz}; the corresponding parity-odd three-current structure in four-dimensional CFT has also been discussed in \cite{CarrilloGonzalez:2026eum}.}
\\
The natural next step would be to test the double-copy relation between the correlators $\langle TTT\rangle$ and $ \langle JJJ\rangle$.
However, before doing so, it is necessary to examine carefully the role of the Schouten identities. As we will discuss, these identities can obscure or make manifest the double-copy structure depending on the tensor basis used to represent the amplitudes.

\subsection{Schouten identities}
The conformal solutions describing the chiral-anomaly interactions were presented in Eqs.~\eqref{eq:avvconf} and \eqref{eq:aaaconf} in terms of specific tensor structures multiplied by the corresponding form factors. This representation, however, is not unique.
By exploiting Schouten identities, the correlators can be rewritten in equivalent forms. 
These identities reflect the fact that not all tensor structures that can be constructed are linearly independent. Such identities arise from the
dimensional degeneracies of tensor structures, given that we are working in $d=4$. In particular, one can write
\begin{equation}
	\begin{aligned}
		0=\varepsilon^{[\alpha_1,\alpha_2,\mu_1,\mu_2,}\bs{p}_1^{\mu_3]} \,{\bs{p}_1}_{ \alpha_1}{\bs{p}_2}_{ \alpha_2}\\
		0=\varepsilon^{[\alpha_1,\alpha_2,\mu_1,\mu_2,}\bs{p}_2^{\mu_3]} \,{\bs{p}_1}_{ \alpha_1}{\bs{p}_2}_{ \alpha_2}
	\end{aligned}
\end{equation}
The expressions above involve the antisymmetrization of five indices. Since we work in four dimensions, at least two of these indices must coincide, implying that the completely antisymmetric combination vanishes identically. These identities can therefore be used to rewrite our conformal solutions.\\
In principle, Schouten identities could be applied directly to Eqs.~\eqref{eq:avvconf} and \eqref{eq:aaaconf}. For our purposes, however, it is more convenient to first contract the free Lorentz indices $\mu_i$ with the corresponding polarisation vectors $\epsilon_i$ and then take the flat-space limit $E\rightarrow0$. In this limit, the identities reduce to
\begin{equation}
	\begin{aligned}
		&{p}_2 \varepsilon^{\bs{p}_1 \bs{\epsilon}_1 \bs{\epsilon}_2 \bs{\epsilon}_3}-{p}_1 \varepsilon^{\bs{p}_2 \bs{\epsilon}_1 \bs{\epsilon}_2 \bs{\epsilon}_3}=\frac{1}{{p}_1}\bigg[({p}_1 \cdot {\epsilon}_3)\varepsilon^{\bs{p}_1\bs{p}_2\bs{\epsilon}_1\bs{\epsilon}_2}-({p}_1 \cdot {\epsilon}_2)\varepsilon^{\bs{p}_1\bs{p}_2\bs{\epsilon}_1\bs{\epsilon}_3}\bigg]+\mathcal{O}(E)\\&
		{p}_2 \varepsilon^{\bs{p}_1 \bs{\epsilon}_1 \bs{\epsilon}_2 \bs{\epsilon}_3}-{p}_1 \varepsilon^{\bs{p}_2 \bs{\epsilon}_1 \bs{\epsilon}_2 \bs{\epsilon}_3}=\frac{1}{{p}_2}\bigg[-({p}_1 \cdot {\epsilon}_3)\varepsilon^{\bs{p}_1\bs{p}_2\bs{\epsilon}_1\bs{\epsilon}_2}+({p}_2 \cdot {\epsilon}_1)\varepsilon^{\bs{p}_1\bs{p}_2\bs{\epsilon}_2\bs{\epsilon}_3}\bigg]+\mathcal{O}(E)
	\end{aligned}
\end{equation}
Note that the use of such identities may hide the Bose symmetric nature of the interaction. 
They allow the flat-space limit of the conformal solution in Eq.~\eqref{eq:avvlimitflat} to be rewritten in several equivalent ways. One may use either identity, or any combination of the two; the particular choice is not relevant for our purposes. Here we adopt the first identity, as an example, obtaining the following representation of the chiral-anomaly correlator
\begin{equation}\label{eq:avvlimitflatdue}
	\lim _{E \rightarrow 0}\frac{E^{{5}/{2}}}{c_{123}^{{1}/{2}}}\, \langle J J J\rangle_{odd} \propto   
	\frac{1}{{p}_1}\bigg[({p}_1 \cdot {\epsilon}_3)\varepsilon^{\bs{p}_1\bs{p}_2\bs{\epsilon}_1\bs{\epsilon}_2}-({p}_1 \cdot {\epsilon}_2)\varepsilon^{\bs{p}_1\bs{p}_2\bs{\epsilon}_1\bs{\epsilon}_3}\bigg]
\end{equation}
This illustrates an important subtlety in the uplift from CFT correlators to flat-space amplitudes: the five-dimensional representation is not unique. 
Tensor structures related by four-dimensional Schouten identities describe the same CFT correlator, but may lead to apparently different representations of the corresponding flat-space amplitude. 
In general, these different five-dimensional expressions need not coincide as fully covariant off-shell extensions; rather, they become equivalent once restricted to the CFT kinematic slice described in Section~\ref{review}. 
In Eq.~\eqref{eq:avvlimitflat}, the chosen tensor basis makes the five-dimensional interpretation of the correlator explicit. 
However, the double-copy structure is not equally transparent in all representations. 
Using the basis of Eq.~\eqref{eq:avvlimitflatdue}, the double-copy relation between this interaction and the conformal-anomaly correlator becomes manifest.

\section{The conformal-anomaly correlator} \label{sec:confanomalcorrel}
In the previous section, we rewrote the correlators associated with the chiral anomaly in a form suitable for establishing the double-copy relations. We now turn to the correlator governed by the conformal anomaly.
The form-factor decomposition of the three-point function of the stress-energy tensor is
\begin{align}
	\left\langle T T T \right\rangle_{even}
	&\nonumber =A_{1}(p_1,p_2,p_3)\left(\epsilon_{1}\cdot p_{2}\,\epsilon_{2}\cdot p_{3}\,\epsilon_{3}\cdot p_{1}\right)^{2}  \\ 
	\nonumber &\quad +\big(A_{2}(p_1,p_2,p_3)\,\epsilon_{1}\cdot\epsilon_{2}\,\epsilon_{1}\cdot p_{2}\,\epsilon_{2}\cdot p_{3}\left(\epsilon_{3}\cdot p_{1}\right)^{2} + \mathrm{cyclic}\big)\\
	\nonumber &\quad +\big(A_{3}(p_1,p_2,p_3)\left(\epsilon_{1}\cdot\epsilon_{2}\right)^{2}\left(p_{1}\cdot\epsilon_{3}\right)^{2}+\mathrm{cyclic}\big)\\
	\nonumber &\quad +\big(A_{4}(p_1,p_2,p_3)\,\epsilon_{1}\cdot\epsilon_{3}\,\epsilon_{2}\cdot\epsilon_{3}\,\epsilon_{1}\cdot p_{2}\,\epsilon_{2}\cdot p_{3}+\mathrm{cyclic}\big)\\
	&\quad +A_{5}(p_1,p_2,p_3)\,\epsilon_{1}\cdot\epsilon_{2}\,\epsilon_{2}\cdot\epsilon_{3}\,\epsilon_{3}\cdot\epsilon_{1}
	\label{tttgeneral} 
\end{align}
where we have contracted the correlator with the polarisation vectors.
The renormalized form factors were obtained in \cite{Bzowski:2017poo} by solving the anomalous conformal ward identities. They are given by
\begin{equation}
	\begin{aligned}\label{A1TTTren}
		A_1 & = c_1 I_{7\{222\}}+\ldots, \\
		A_2 & =  2 \Big[b_1 + b_2 - 2\, c_1 \, p_3 \frac{\partial}{\partial p_3} \Big] I^{\text{(fin)}}_{5\{222\}}+\ldots,\\
		A_3 & = 2 \Big[ 2 b_2 - (b_1 + b_2 + c_1) p_3 \frac{\partial}{\partial p_3} + c_1 p_3^2 \frac{\partial^2}{\partial p_3^2} \Big] I_{3\{222\}}^{\text{(fin)}}+\ldots,\\
		A_4 & = 4 \Big[ b_2-b_1 + (b_1+b_2) p_3 \frac{\partial}{\partial p_3}+2 c_1 \Big( 8 - 4 \sum_{j=1}^3 p_j \frac{\partial}{\partial p_j} + p_1 p_2 \frac{\partial^2}{\partial p_1 \partial p_2} \Big) \Big] I_{3\{222\}}^{\text{(fin)}}+\ldots,\\
		A_5 & = 2(b_1+b_2) \Big[ 32 - 8 \sum_{j=1}^3 p_j \frac{\partial}{\partial p_j} + 2 \sum_{i<j} p_i p_j \frac{\partial^2}{\partial p_i \partial p_j} \Big] I_{1\{222\}}^{\text{(fin)}}  \\&\quad 
		- 8 c_1 p_1^3 p_2^3 p_3^3 \frac{\partial^3}{\partial p_1 \partial p_2 \partial p_3} I_{1\{000\}} 
		+\ldots,
	\end{aligned}
\end{equation}
where the ellipses denote terms that are non-singular in the flat space limit. The coefficients 
$b_1$ and $b_2$ are those entering the trace anomaly \eqref{eq:anomtrac}, while $c_1$ is an independent coefficient associated with the homogeneous solution of the conformal Ward identities. The 3K integral $I_{7\{222\}}$ and the finite integrals $ I_{5\{222\}}^{\text{(fin)}}$, $I_{3\{222\}}^{\text{(fin)}}$ and $I_{1\{222\}}^{\text{(fin)}}$ can all be expressed as derivatives of the master integral $I_{1\{000\}}$, see Appendix \ref{app3kexpression}. Up to an overall constant, this master integral coincides with the one-loop triangle integral $C_0(p_1,p_2,p_3)$.
\\
The flat-space limits of the correlator then take the form \cite{Lipstein:2019mpu}
\begin{equation}
	\begin{aligned} \label{4dTTTflat1}
		&\lim_{E\rightarrow0}\frac{E^{13/2}}{c_{123}^{3/2}}\left\langle TTT\right\rangle_{even} \propto c_1 \mathcal{A}_{W^{3}}, \\
		&\lim_{E\rightarrow0}\left.\frac{E^{9/2}}{c_{123}^{3/2}}\left\langle TTT\right\rangle_{even} \right|_{c_{1}=0}\propto (b_1+b_2) \mathcal{A}_{\phi R^{2}}^{222}, \\
		&\lim_{E\rightarrow0}\left.\frac{E^{5/2}}{c_{123}^{3/2}}\left\langle TTT\right\rangle_{even} \right|_{c_{1}=0, \, \, b_1+b_2=0}\propto b_2\, \mathcal{A}_{EG}.
	\end{aligned}
\end{equation}
The first contribution is controlled by the coefficient $c_1$, which parametrizes the non-anomalous homogeneous part of the correlator and gives rise to the higher-derivative $W^3$ interaction in the flat-space limit. Conversely, the coefficients multiplying the $\phi R^2$ and Einstein-gravity amplitudes are determined by the trace-anomaly coefficients. This connection is natural from a holographic perspective, since these anomaly coefficients are determined by bulk gravitational interactions   \cite{Henningson:1998gx,Nojiri:1999mh,Bugini:2016nvn}. In the case of Einstein gravity, for instance, one has $b_1+b_2=0$, and the contribution proportional to $\mathcal{A}_{\phi R^2}^{222}$ consequently vanishes. Related on-shell kinematic structures for higher-derivative
graviton amplitudes have also been discussed in
\cite{Chowdhury:2019kaq}.

\section{Double copy between chiral- and conformal-anomaly correlators}\label{sec:dcopyavvttt}

In this section, we establish a connection between the chiral- and conformal-anomaly correlators through a double-copy relation. Since each correlator admits a decomposition into parity-even and parity-odd sectors, the possible double-copy relations can be schematically written as
\begin{equation}\label{eq:tttavvdoublecopy}
	\begin{aligned}
		&  \langle TTT\rangle_{even} \propto \langle JJJ\rangle_{even} \, \langle JJJ\rangle_{even},\\
		&    \langle TTT\rangle_{even} \propto  \langle JJJ\rangle_{odd}\, \langle JJJ\rangle_{odd} ,\\
		&  \langle TTT\rangle_{odd} \, \, \, \propto \langle JJJ\rangle_{even} \, \langle JJJ\rangle_{odd} 
	\end{aligned}
\end{equation}
The first of these relations was established in \cite{Farrow:2018yni,Lipstein:2019mpu}. The second is the main focus of the present section.
The third relation requires a separate discussion. For it to be nontrivial, the parity-odd stress-tensor correlator $\langle TTT\rangle_{odd}$ must be nonvanishing. The conformal analysis of this correlator has so far been carried out only in position space, where it was found to vanish \cite{Stanev:2012nq,Zhiboedov:2012bm}. This result is closely related to the assumption that parity-odd conformal anomalies are absent in \eqref{eq:anomtrac}. The status of such anomalies, however, remains under debate, as perturbative and nonperturbative analyses have led to apparently conflicting conclusions \cite{Abdallah:2023cdw,Bonora:2022izj,Bonora:2026mff}. At present, no nonvanishing momentum-space conformal solution for $\langle TTT\rangle_{odd}$ is known. We therefore restrict our attention to the second relation in \eqref{eq:tttavvdoublecopy}.\\
We recall that the flat-space limit of the parity-odd current correlator is
\begin{equation}
	\lim _{E \rightarrow 0}\frac{E^{{5}/{2}}}{c_{123}^{{1}/{2}}}\, \langle J J J\rangle_{odd} \propto  
	\frac{ a_1 }{{p}_1}\bigg[({p}_1 \cdot {\epsilon}_3)\varepsilon^{\bs{p}_1\bs{p}_2\bs{\epsilon}_1\bs{\epsilon}_2}-({p}_1 \cdot {\epsilon}_2)\varepsilon^{\bs{p}_1\bs{p}_2\bs{\epsilon}_1\bs{\epsilon}_3}\bigg]
\end{equation}
while the flat-space limit of the stress-tensor three-point function, restricted to the sector with $c_1=0$, is
 \begin{equation}
 	\lim_{E\rightarrow0}\left.\frac{E^{9/2}}{c_{123}^{3/2}}\left\langle TTT\right\rangle_{even} \right|_{c_{1}=0}\propto (b_1+b_2) \, \mathcal{A}_{\phi R^{2}}^{222}
 \end{equation}
The coefficient $c_1$ is unrelated to the trace-anomaly coefficients, whereas the remaining contribution is entirely controlled by the combination $b_1+b_2$. Imposing $c_1=0$ therefore isolates the component of the stress-tensor correlator that is governed exclusively by the conformal anomaly.\\
Squaring the parity-odd current correlator, we obtain
 \begin{equation}
 	\lim _{E \rightarrow 0}E^5 \, \langle J J J\rangle_{odd} \, \langle J J J\rangle_{odd} \propto\left. \lim _{E \rightarrow 0}\frac{E^{9/2}}{c_{123}^{1/2}}\left\langle TTT\right\rangle_{even} \right|_{c_{1}=0}
 \end{equation}
This relation establishes a direct connection between the chiral and conformal anomalies. As shown in \cite{Coriano:2023hts}, the parity-odd correlator $\langle JJJ\rangle_{odd}$ is generated entirely by the chiral anomaly and vanishes when the anomaly coefficient is set to zero. Similarly, after imposing $c_1=0$, the surviving contribution to $\langle TTT\rangle_{even}$ is completely determined by the conformal-anomaly coefficients. The double-copy relation therefore maps a correlator generated solely by the chiral anomaly into a stress-tensor correlator whose surviving contribution is governed entirely by the conformal anomaly.
\\
It is important to note that the relations established here are proportionality relations rather than exact equalities, as the latter would require a direct connection between the corresponding anomaly coefficients. In this regard, it is worth emphasizing that, although the coefficients of the chiral and conformal anomalies are generally independent, they can be related in specific classes of theories. A notable example is provided by superconformal field theories, where supersymmetry establishes exact relations between the conformal anomaly coefficients and the chiral anomalies of the $U(1)_R$ current \cite{Anselmi:1997am,Intriligator:2003jj}. \\
The approach developed in this work does not allow us to establish a direct relation between these coefficients. Indeed, the solutions of the conformal Ward identities determine the corresponding correlators only up to overall constants, which remain theory-dependent. The double-copy relations derived here establish a correspondence between the tensorial structures and their momentum dependence, rather than between their overall normalizations. Consequently, the anomaly coefficients entering the chiral and conformal correlators remain independent within our framework. It would be interesting to investigate in greater depth how additional constraints or complementary approaches could establish further relations between these coefficients.

 \section{Conclusions}\label{sec:conclusions}

In this work, we have shown that the relation between chiral and conformal anomalies admits a natural interpretation in terms of a double-copy structure within four-dimensional conformal field theory. What initially appears as a factorization of local anomaly densities extends to complete momentum-space correlators and, through their flat-space limits, to higher-dimensional amplitudes.
 \\
 We first analysed three-point correlators involving two conserved currents or stress tensors and a marginal scalar operator. These correlators provide a simple setting in which the relation between gauge and gravitational structures can be identified directly at the level of conformal correlators. By decomposing the correlators into parity-even and parity-odd sectors, we showed that the structures appearing in the $\langle JJO\rangle$ correlators combine to reproduce the corresponding structures of the $\langle TTO\rangle$ correlators through double-copy-type relations. 
 \\
  We then turned to the correlators governed by the chiral anomaly. The anomalous three-current correlators $\langle J_VJ_VJ_A\rangle$ and $\langle J_AJ_AJ_A\rangle$ are completely fixed by the combination of conformal symmetry and the anomalous chiral Ward identities. Although these correlators involve different combinations of vector and axial currents, their flat-space limits are governed by the same parity-odd five-dimensional amplitude. This universal structure is generated by a Chern--Simons interaction and reflects the topological information encoded in the anomaly.
 \\
 An important aspect of our analysis is that the higher-dimensional uplift of a four-dimensional correlator is not unique. Different tensor representations of the same CFT correlator can be related through Schouten identities in $4d$. Accounting for this freedom is crucial for revealing double-copy structures that can remain obscured in a particular tensor basis.
 \\
On the gravitational side, we examine the known flat-space limit of the parity-even $\langle TTT\rangle$ correlator and isolate the sector entirely determined by the conformal anomaly. This is achieved by removing the contributions associated with the homogeneous solutions of the conformal Ward identities. We then show that the resulting anomaly-controlled gravitational structure is reproduced by the double copy of the parity-odd current amplitude associated with the chiral anomaly. \\
This establishes a direct correspondence between the chiral and conformal anomaly sectors, extending the relation between anomaly densities to complete conformal correlators and their higher-dimensional amplitude limits.
 More broadly, anomaly-induced correlators provide a promising framework for investigating how quantum effects can reveal unexpected connections between gauge and gravitational theories. Several directions remain open, including the extension to higher-point functions and the study of possible mixed-parity gravitational structures.
 \\
 \\
\centerline{\bf Acknowledgements} 
 This work is partially supported by INFN, inziativa specifica {\em QG-sky}, by the the grant PRIN 2022BP52A MUR "The Holographic Universe for all Lambdas" Lecce-Naples, and by the European Union, Next Generation EU, PNRR project "National Centre for HPC, Big Data and Quantum Computing", project code CN00000013.

 \appendix
 \section{Functional derivatives of anomaly densities} \label{appfuncder}

 In this appendix, we report the expressions for the functional derivatives of various gauge and gravitational densities that may appear in anomalous Ward identities, such as \eqref{eq:anomTTTWI} and \eqref{eq:anomJJJWI}. We omit overall numerical factors, as they can be absorbed into a redefinition of the anomaly coefficients.
 The relevant parity-even structures are given by  
 \begin{equation}
	\mathcal{A}^{\mu_1\mu_2}_{F^2}=\left(\bs{p}_1 \cdot \bs{p}_2\right) \delta_{\alpha_1\alpha_2}-{\bs{p}_1}_{\alpha_2}{\bs{p}_2}_{\alpha_1}
\end{equation}
and
  \begin{equation}\label{eq:dfunzeuler}
 	\begin{aligned}
 		\mathcal{A}^{\mu_1\nu_1\mu_2\nu_2}_{E_4}&= 
 		 \delta^{\mu_1 \nu_2} \delta^{\mu_2 \nu_1}(\bs{p}_1 \cdot \bs{p}_2)^2-2 \delta^{\mu_1 \nu_1} \delta^{\mu_2 \nu_2}(\bs{p}_1 \cdot \bs{p}_2)^2+\delta^{\mu_1 \mu_2} \delta^{\nu_1 \nu_2}(\bs{p}_1 \cdot \bs{p}_2)^2-\bs{p}_2^{\mu_1} \bs{p}_1^{\mu_2} \delta^{\nu_1 \nu_2}(\bs{p}_1 \cdot \bs{p}_2)\\
 		& -\bs{p}_1^{\mu_1} \bs{p}_2^{\mu_2} \delta^{\nu_1 \nu_2}(\bs{p}_1 \cdot \bs{p}_2)+2 \bs{p}_2^{\mu_1} \delta^{\mu_2 \nu_2} \bs{p}_1^{\nu_1}(\bs{p}_1 \cdot \bs{p}_2)-\delta^{\mu_1 \nu_2} \bs{p}_2^{\mu_2} \bs{p}_1^{\nu_1}(\bs{p}_1 \cdot \bs{p}_2)+2 \bs{p}_1^{\mu_1} \delta^{\mu_2 \nu_2} \bs{p}_2^{\nu_1}(\bs{p}_1 \cdot \bs{p}_2) \\
 		&- \delta^{\mu_1 \nu_2} \bs{p}_1^{\mu_2} \bs{p}_2^{\nu_1}(\bs{p}_1 \cdot \bs{p}_2)-\bs{p}_2^{\mu_1} \delta^{\mu_2 \nu_1} \bs{p}_1^{\nu_2}(\bs{p}_1 \cdot \bs{p}_2)+2 \delta^{\mu_1 \nu_1} \bs{p}_2^{\mu_2} \bs{p}_1^{\nu_2}(\bs{p}_1 \cdot \bs{p}_2)-\delta^{\mu_1 \mu_2} \bs{p}_2^{\nu_1} \bs{p}_1^{\nu_2}(\bs{p}_1 \cdot \bs{p}_2)\\
 		&- \bs{p}_1^{\mu_1} \delta^{\mu_2 \nu_1} \bs{p}_2^{\nu_2}(\bs{p}_1 \cdot \bs{p}_2)+2 \delta^{\mu_1 \nu_1} \bs{p}_1^{\mu_2} \bs{p}_2^{\nu_2}(\bs{p}_1 \cdot \bs{p}_2)-\delta^{\mu_1 \mu_2} \bs{p}_1^{\nu_1} \bs{p}_2^{\nu_2}(\bs{p}_1 \cdot \bs{p}_2)-\bs{p}_2^{\mu_1} \bs{p}_2^{\mu_2} \bs{p}_1^{\nu_1} \bs{p}_1^{\nu_2} \\
 		&+ 2 \bs{p}_2^{\mu_1} \bs{p}_1^{\mu_2} \bs{p}_2^{\nu_1} \bs{p}_1^{\nu_2}-\bs{p}_1^{\mu_1} \bs{p}_2^{\mu_2} \bs{p}_2^{\nu_1} \bs{p}_1^{\nu_2}-\bs{p}_2^{\mu_1} \bs{p}_1^{\mu_2} \bs{p}_1^{\nu_1} \bs{p}_2^{\nu_2}+2 \bs{p}_1^{\mu_1} \bs{p}_2^{\mu_2} \bs{p}_1^{\nu_1} \bs{p}_2^{\nu_2}-\bs{p}_1^{\mu_1} \bs{p}_1^{\mu_2} \bs{p}_2^{\nu_1} \bs{p}_2^{\nu_2} \\
 		&+ \bs{p}_2^{\mu_1} \bs{p}_2^{\mu_2} \delta^{\nu_1 \nu_2} \bs{p}_1^2-2 \bs{p}_2^{\mu_1} \delta^{\mu_2 \nu_2} \bs{p}_2^{\nu_1} \bs{p}_1^2+\delta^{\mu_1 \nu_2} \bs{p}_2^{\mu_2} \bs{p}_2^{\nu_1} \bs{p}_1^2+\bs{p}_2^{\mu_1} \delta^{\mu_2 \nu_1} \bs{p}_2^{\nu_2} \bs{p}_1^2-2 \delta^{\mu_1 \nu_1} \bs{p}_2^{\mu_2} \bs{p}_2^{\nu_2} \bs{p}_1^2 \\
 		&+ \delta^{\mu_1 \mu_2} \bs{p}_2^{\nu_1} \bs{p}_2^{\nu_2} \bs{p}_1^2+\bs{p}_1^{\mu_1} \bs{p}_1^{\mu_2} \delta^{\nu_1 \nu_2} \bs{p}_2^2-2 \bs{p}_1^{\mu_1} \delta^{\mu_2 \nu_2} \bs{p}_1^{\nu_1} \bs{p}_2^2+\delta^{\mu_1 \nu_2} \bs{p}_1^{\mu_2} \bs{p}_1^{\nu_1} \bs{p}_2^2+\bs{p}_1^{\mu_1} \delta^{\mu_2 \nu_1} \bs{p}_1^{\nu_2} \bs{p}_2^2 \\
 		&- 2 \delta^{\mu_1 \nu_1} \bs{p}_1^{\mu_2} \bs{p}_1^{\nu_2} \bs{p}_2^2+\delta^{\mu_1 \mu_2} \bs{p}_1^{\nu_1} \bs{p}_1^{\nu_2} \bs{p}_2^2-\delta^{\mu_1 \nu_2} \delta^{\mu_2 \nu_1} \bs{p}_1^2 \bs{p}_2^2+2 \delta^{\mu_1 \nu_1} \delta^{\mu_2 \nu_2} \bs{p}_1^2 \bs{p}_2^2-\delta^{\mu_1 \mu_2} \delta^{\nu_1 \nu_2} \bs{p}_1^2 \bs{p}_2^2
 	\end{aligned}
 \end{equation}
while the parity-odd structures are
\begin{equation}
	\mathcal{A}^{\mu_1\mu_2}_{F\tilde{F}}=\epsilon^{\mu_{1}\mu_{2}p_1p_2}
\end{equation}
and
 \begin{equation}
 	\mathcal{A}^{\mu_1\nu_1\mu_2\nu_2}_{R\widetilde{R}}
 	=	\varepsilon^{\nu_1 \nu_2 \bs{p}_1 \bs{p}_2}\bigg[\left(\bs{p}_1 \cdot \bs{p}_2\right) \delta^{\mu_1 \mu_2}-\bs{p}_1^{\mu_2} \bs{p}_2^{\mu_1}\bigg]+\left(\mu_1 \leftrightarrow \nu_1\right)+\left(\mu_2 \leftrightarrow \nu_2\right)+\binom{\mu_1 \leftrightarrow \nu_1}{\mu_2 \leftrightarrow \nu_2} 
 \end{equation}
 As expected, all the structures listed above are transverse with respect to the momenta. Moreover, $\mathcal{A}^{\mu_1\nu_1\mu_2\nu_2}_{R\widetilde{R}}$ is also traceless, whereas $\mathcal{A}^{\mu_1\nu_1\mu_2\nu_2}_{E_4}$ is not.
 
\section{The conformal solution for the $\langle JJO\rangle_{even}$ and $\langle TTO\rangle_{even}$}\label{appJJOTTO}
 
In this appendix, we collect the conformal solutions for the $\langle JJO\rangle_{{even}}$ and $\langle TTO\rangle_{{even}}$ correlators derived in \cite{Bzowski:2018fql}. We employ the following projectors
 \begin{equation}
 	\pi_\alpha^\mu(\bs{p})=\delta_\alpha^\mu-\frac{\bs{p}^\mu \bs{p}_\alpha}{\bs{p}^2}, \qquad\qquad 
 		\Pi_{\alpha \beta}^{\mu \nu}=\frac{1}{2}\left(\pi_\alpha^\mu \pi_\beta^\nu+\pi_\beta^\mu \pi_\alpha^\nu\right)-\frac{1}{d-1} \pi^{\mu \nu} \pi_{\alpha \beta}.
 \end{equation}
The first projector is used in the decomposition of the $\langle JJO\rangle_{{even}}$ correlator, while the second projects onto the transverse-traceless part of $\langle TTO\rangle_{{even}}$.
\\
The general decomposition of $\langle JJO\rangle_{{even}}$ can be written in terms of two form factors
 \begin{equation}
 	\left\langle J^{\mu_1 }\left(\boldsymbol{p}_1\right) J^{\mu_2 }\left(\boldsymbol{p}_2\right) {O}\left(\boldsymbol{p}_3\right)\right\rangle_{{even}}=\pi_{\alpha_1}^{\mu_1}\left(\boldsymbol{p}_1\right) \pi_{\alpha_2}^{\mu_2}\left(\boldsymbol{p}_2\right)\left[A_1 \bs{p}_2^{\alpha_1} \bs{p}_3^{\alpha_2}+A_2 \delta^{\alpha_1 \alpha_2}\right]
 \end{equation}
Imposing the conformal constraints, for the case $d=\Delta_3=4$, one finds
 \begin{equation}
 	\begin{aligned}
 		A_1 = & c_1 \left(2-p_3 \frac{\partial}{\partial p_3}\right) I_{2\{111\}}^{(\text {fin})}-\frac{1}{3} c_1 \left(\ln \frac{p_1^2}{\mu^2}+\ln \frac{p_2^2}{\mu^2}+\ln \frac{p_3^2}{\mu^2}\right), \\
 		A_2 = & c_1 \, p_3^2 I_{2\{111\}}^{(\text {fin})}+\frac{1}{6} c_1 \left[\left(3 p_1^2-p_3^2\right) \ln \frac{p_1^2}{\mu^2}+\left(3 p_2^2-p_3^2\right) \ln \frac{p_2^2}{\mu^2}-p_3^2 \ln \frac{p_3^2}{\mu^2}+\frac{p_1^2+p_2^2+5p_3^2}{2} \right]
 	\end{aligned}
 \end{equation}
 where the finit integral $I_{2\{111\}}^{(\text {fin})}$ is given in Appendix \ref{app3kexpression}. \\
The above solution can be supplemented by additional scheme-dependent finite contributions arising from the counterterm action
\begin{equation}
	d_1 \int \mathrm{d}^d {x} \sqrt{g}  \, \phi \, F^2
\end{equation}
This counterterm generates the contributions
\begin{equation}
	\begin{aligned}
		A_1^{\mathrm{ct} } & =  d_1 , \\
		A_2^{\mathrm{ct} } & =-d_1 \frac{p_1^2+p_2^2-p_3^2}{2}.
	\end{aligned}
\end{equation}
These contributions correspond to the term
$\mathcal{A}^{\mu_1\mu_2}_{\phi F^2}$ in Eq.~\eqref{eq:jjoeven}.\\
On the other hand, the general decomposition of the $\langle TTO\rangle_{{even}}$ correlator is more involved, since it contains both anomalous trace contributions and transverse-traceless terms
\begin{equation}\label{eq:ttodecompgeneral}
	\begin{gathered}
		\left\langle T^{\mu_1 \nu_1}\left(\boldsymbol{p}_1\right) T^{\mu_2 \nu_2}\left(\boldsymbol{p}_2\right) {O}\left(\boldsymbol{p}_3\right)\right\rangle_{{even}}=\Pi^{\mu_1 \nu_1}_{\alpha_1 \beta_1}\left(\boldsymbol{p}_1\right) \Pi^{\mu_2 \nu_2}_{\alpha_2 \beta_2}\left(\boldsymbol{p}_2\right)X^{\alpha_1\beta_{1}\alpha_2\beta_{2}}+\frac{1}{d-1} \pi^{\mu_1 \nu_1}\left(\boldsymbol{p}_1\right) \mathcal{A}^{\mu_2 \nu_2} \\
		+\frac{1}{d-1} \pi^{\mu_2 \nu_2}\left(\boldsymbol{p}_2\right) \mathcal{A}^{\mu_1 \nu_1}\left(\boldsymbol{p}_1 \leftrightarrow \boldsymbol{p}_2\right)-\frac{1}{(d-1)^2} \pi^{\mu_1 \nu_1}\left(\boldsymbol{p}_1\right) \pi^{\mu_2 \nu_2}\left(\boldsymbol{p}_2\right) \delta^{\alpha \beta} \mathcal{A}_{\alpha \beta}.
	\end{gathered}
\end{equation}
where the anomalous contribution is defined as
\begin{equation}\label{eq:tracetto}
	\mathcal{A}^{\mu_2 \nu_2} =\delta_{\mu_1\nu_1}\left\langle T^{\mu_1 \nu_1}\left(\boldsymbol{p}_1\right) T^{\mu_2 \nu_2}\left(\boldsymbol{p}_2\right) {O}\left(\boldsymbol{p}_3\right)\right\rangle_{{even}}
\end{equation}
We have also introduced the rank-four tensor $X^{\alpha_1\beta_1\alpha_2\beta_2}$, which parametrizes the transverse-traceless part and admits the following general decomposition in terms of form factors
 \begin{equation}
 	\begin{aligned}
 		 X^{\alpha_1\beta_{1}\alpha_2\beta_{2}}=A_1 \bs{p}_2^{\alpha_1} \bs{p}_2^{\beta_1} \bs{p}_3^{\alpha_2} \bs{p}_3^{\beta_2}+A_2 \delta^{\alpha_1 \alpha_2} \bs{p}_2^{\beta_1} \bs{p}_3^{\beta_2}+A_3 \delta^{\alpha_1 \alpha_2} \delta^{\beta_1 \beta_2} 
 	\end{aligned}
 \end{equation}
 The corresponding form factors are again completely determined by the conformal Ward identities. For $d=\Delta_3=4$, one finds
 \begin{equation}
 	\begin{aligned}
 		A_1= & c_1\left(2-p_1 \frac{\partial}{\partial p_1}\right)\left(2-p_2 \frac{\partial}{\partial p_2}\right)\left(2-p_3 \frac{\partial}{\partial p_3}\right) I_{2\{111\}}^{(\text{fin})} 
 		 -\frac{4}{3} c_1\left[\ln \frac{p_1^2}{\mu^2}+\ln \frac{p_2^2}{\mu^2}+\ln \frac{p_3^2}{\mu^2}\right]-8 c_1 , \\
 		A_2= & 4 c_1\left(1-p_3 \frac{\partial}{\partial p_3}\right) I_{3\{222\}}^{(\text {fin})} 
 		 +2 c_1\left[\left(p_1^2+p_2^2\right) \ln \frac{p_3^2}{\mu^2}+\left(p_1^2-p_3^2\right) \ln \frac{p_2^2}{\mu^2}+\left(p_2^2-p_3^2\right) \ln \frac{p_1^2}{\mu^2}\right]  +4 c_1 p_3^2, \\
 		A_3= & 2 c_1 p_3^2 I_{3\{222\}}^{(\text{fin})}+c_1\bigg[\left(p_2^2 p_3^2-p_1^4+p_3^4\right) \ln \frac{p_1^2}{\mu^2}+\left(p_1^2 p_3^2-p_2^4+p_3^4\right) \ln \frac{p_2^2}{\mu^2}+p_3^2\left(p_1^2+p_2^2-3 p_3^2\right) \ln \frac{p_3^2}{\mu^2} \\
 		&  -2\left(p_1^4+p_2^4-p_1^2 p_2^2+2 p_3^4\right)\bigg] 
 		 ,
 	\end{aligned}
 \end{equation}
See Appendix \ref{app3kexpression}, for the expression of the 3K integrals above. 
  Furthermore, the counterterm action
  \begin{equation}
  	\int \mathrm{d}^d {x} \sqrt{g}  \phi \left(d_1 E_4+d_2 W_4^2\right)
  \end{equation}
generates the additional finite contributions
  \begin{equation}
  	\begin{aligned}
  		A_1^{\mathrm{ct} } & =8\left(d_1 +d_2\right)  , \\
  		A_2^{\mathrm{ct} } & =-8\left(d_1 +d_2 \right)\left(p_1^2+p_2^2-p_3^2\right) , \\
  		A_3^{\mathrm{ct} } & = 4 \, d_2\,  p_1^2 p_2^2-2\left(d_1 +d_2 \right) J^2  .
  	\end{aligned}
  \end{equation}
The counterterm action also gives rise to trace contributions that enter the anomalous sector of the  correlator, as shown in Eq.~\eqref{eq:tracetto}. In particular, the $\phi E_4$ interaction generates the following anomalous contribution
  \begin{equation}
 	\mathcal{A}^{\mu_2 \nu_2} =d_1 \bigg[\frac{1}{4}J^2\delta^{\mu_2\nu_2}+\frac{(p_3^2-p_1^2-p_2^2)}{2}\left(\bs{p}_2^{\mu_2}\bs{p}_1^{\nu_2}+\bs{p}_1^{\mu_2}\bs{p}_2^{\nu_2}\right)
 	-p_1^2 \, \bs{p}_2^{\mu_2}\bs{p}_2^{\nu_2}-p_2^2\, \bs{p}_1^{\mu_2}\bs{p}_1^{\nu_2}\bigg]
 \end{equation}
This contribution must be included in the decomposition given in Eq.~\eqref{eq:ttodecompgeneral}. Conversely, the Weyl term $\phi W^2_4$ does not give rise to any anomalous contribution.\\
The terms parametrized by the coefficient $d_1$ in both the transverse-traceless and trace sectors of the $\langle TTO\rangle_{{even}}$ correlator correspond to the contribution
$\mathcal{A}^{\mu_1\nu_1\mu_2\nu_2}_{\phi E_4}$ appearing in Eq.~\eqref{eq:ttoeven}.
The complete expression for $\mathcal{A}^{\mu_1\nu_1\mu_2\nu_2}_{\phi E_4}$, including both transverse-traceless and trace components, is also reported in Eq.~\eqref{eq:dfunzeuler}, since it coincides with the Euler-density contribution
 $\mathcal{A}^{\mu_1\nu_1\mu_2\nu_2}_{E_4}$ by definition.

 \section{3K integrals and the master integral}\label{app3kexpression}
In this appendix, we collect the expressions for the 3K integrals appearing throughout the paper. In particular, the integrals relevant for our analysis can be expressed in terms of the master integral $I_{1\{000\}}$ and its derivatives as
 \begin{equation}
 	\begin{aligned}
 		 	I_{3\{101\}}&=p_1 p_3 \frac{\partial^2}{\partial p_1 \partial p_3} I_{1\{0,0,0\}},\\
 		I_{3\{011\}}&=p_2 p_3 \frac{\partial^2}{\partial p_2 \partial p_3} I_{1\{0,0,0\}}
 		,\\
 		I_{7\{222\}} & =-\left(2-p_1 \frac{\partial}{\partial p_1}\right)\left(2-p_2 \frac{\partial}{\partial p_2}\right)\left(2-p_3 \frac{\partial}{\partial p_3}\right) p_1 p_2 p_3 \frac{\partial^3}{\partial p_1 \partial p_2 \partial p_3} I_{1\{000\}}, \\
 		I_{5\{222\}}^{(\text {fin})} & =\left(2-p_1 \frac{\partial}{\partial p_1}\right)\left(2-p_2 \frac{\partial}{\partial p_2}\right)\left(2-p_3 \frac{\partial}{\partial p_3}\right) I_{2\{111\}}^{(\text {fin})}, \\
 		I_{3\{222\}}^{(\text {fin})} & =\left(2-p_1 \frac{\partial}{\partial p_1}\right)\left(2-p_2 \frac{\partial}{\partial p_2}\right)\left(2-p_3 \frac{\partial}{\partial p_3}\right)\left(\frac{1}{4} J^2 I_{1\{000\}}\right), \\
 		I_{1\{222\}}^{(\text {fin})} & =\left[p_1^2 p_2^2 p_3^2-\frac{1}{4} J^2\left(p_1^2+p_2^2+p_3^2\right)\right] I_{1\{000\}},
 	\end{aligned}
 \end{equation}
The finite part of the integral $I_{2\{111\}}$ appearing above is given by
  \begin{equation}
 	\begin{aligned}
 		I_{2\{111\}}^{(\text {fin})}=-\frac{4 p_1^2 p_2^2 p_3^2}{J^2} & I_{1\{000\}}-\frac{1}{6 J^2}\left[p_1^2\left(p_2^2+p_3^2-p_1^2\right) \ln \left(\frac{p_1^4}{p_2^2 p_3^2}\right)\right. \\
 		& \left.\quad+p_2^2\left(p_1^2+p_3^2-p_2^2\right) \ln \left(\frac{p_2^4}{p_1^2 p_3^2}\right)+p_3^2\left(p_1^2+p_2^2-p_3^2\right) \ln \left(\frac{p_3^4}{p_1^2 p_2^2}\right)\right]
 	\end{aligned}
 \end{equation} 
The master integral $I_{1\{000\}}$ can be expressed in terms of dilogarithms as
 \begin{equation}
 	I_{1\{000\}}=\frac{1}{2 \sqrt{-J^2}}\left[\frac{\pi^2}{6}-2 \ln \frac{p_1}{p_3} \ln \frac{p_2}{p_3}+\ln X \ln Y-\operatorname{Li}_2 X-\operatorname{Li}_2 Y\right]
 \end{equation}
 where
 \begin{equation}
 	X=\frac{-p_1^2+p_2^2+p_3^2-\sqrt{-J^2}}{2 p_3^2}, \quad Y=\frac{-p_2^2+p_1^2+p_3^2-\sqrt{-J^2}}{2 p_3^2} .
 \end{equation}
Up to an overall normalization factor, this master integral coincides with the one-loop triangle integral $C_0(p_1,p_2,p_3)$.

\providecommand{\href}[2]{#2}\begingroup\raggedright\endgroup

\end{document}